\documentclass[twocolumn,showpacs,preprintnumbers,amsmath,amssymb,prl]{revtex4-1}
\usepackage{mathrsfs}
\usepackage{graphicx}
\usepackage{dcolumn}
\usepackage{bm}
\usepackage{amsmath}
\usepackage{amsfonts}
\begin{document}

\title{Quantum Spin Hall and Quantum Anomalous Hall States Realized in Junction Quantum Wells}

\author{Haijun Zhang, Yong Xu, Jing Wang and Shou-Cheng Zhang}
\affiliation{ Department of Physics, McCullough Building, Stanford University, Stanford, CA 94305-4045, USA}

\begin{abstract}
Both quantum spin Hall and quantum anomalous Hall states are novel states of quantum matter with promising applications. We propose junction quantum wells comprising II-VI, III-V or IV semiconductors as a large class of new materials realizing the quantum spin Hall state. Especially, we find that the bulk band gap for the quantum spin Hall state can be as large as 0.1 eV. Further more, magnetic doping would induce the ferromagnetism in these junction quantum wells due to band edge singularities in the band-inversion regime and to realize the quantum anomalous Hall state.
\end{abstract}

\date{\today}


\maketitle

{\bf Introduction.} Topological states are new states of quantum matter with topologically protected gapless boundary states.\cite{qi2010a,Moore2010,Hasan2010,qi2011} In two dimensions, both the quantum spin Hall (QSH) and quantum anomalous Hall (QAH) states have topologically protected edge states on the boundary, where the electron backscattering is forbidden, offering a promising way for the application of electronic devices without dissipation. The QSH state was theoretically predicted and experimentally observed in HgTe/CdTe quantum well (QW),\cite{bernevig2006d,koenig2007} and subsequently in the type-II InAs/GaSb QW.\cite{liu2008a,knez2011} Soon later, the QAH state was also predicted and observed on Cr-doped Bi$_2$Te$_3$ thin films\cite{Yu2010,Chang2013}. Unfortunately, these effects have only been observed at low temperatures due to the small bulk band gap. Here we propose a large class of junction quantum wells, comprising II-VI, III-V or IV semiconductors, for realizing both QSH and QAH effects, possibly at room temperature.

In this Letter, we first generally discuss the principle of obtaining the topologically nontrivial electronic structure in junction quantum wells.  Then we take InSb-based junction quantum wells as an illustrative example to demonstrate how to design both QSH and QAH states, and especially we find that the bulk band gap of the QSH state in junction quantum wells can be as large as 0.1 eV. In addition, a bias voltage in junction quantum wells is proposed to assist to turn on or off QSH (QAH) states, which provides novel functionality for the application of next-generation electronic devices.

\begin{figure}[t]
\begin{center}
\includegraphics[width=2.0in,clip=true,angle=-90]{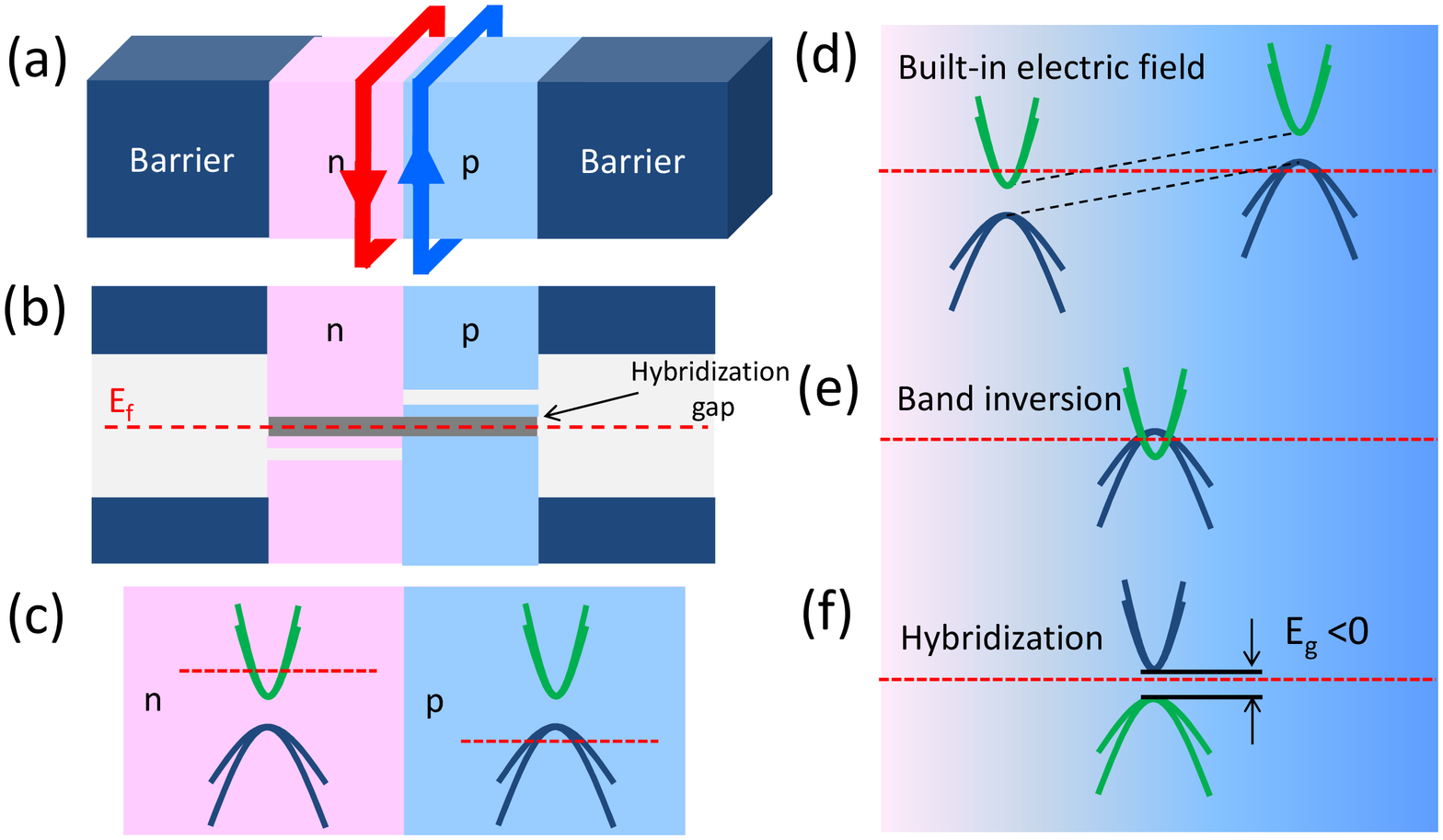}
\end{center}
\caption{(color online). (a) The schematic of a junction quantum well (JQW). The red and bule curves with the arrows are for the topologically protected edge states of the QSH state. (b) The schematic of electronic structure of a JQW. After the bottom of $n$-region conduction bands overlaps the top of $p$-region valence bands, an hybridized band gap could appear within the depletion region. The dashed red lines are for chemical potentials. (c) The schematic of the band structure of a JQW before the charge transfer between the $p$- and $n$- regions. (d) An electric field is consequentially built up to align the chemical potentials of $p$- and $n$- regions after the charge transfer is completed. (e) A band inversion occurs in the case of heavy doping, (e.g. $\sim$0.1 nm$^{-3}$). (f) The hybridization of wavefunctions opens a bulk band gap. E$_g$ is defined as the band gap at $\Gamma$.
}\label{fig1}
\end{figure}

\begin{figure*}[t]
\begin{center}
\includegraphics[width=4.5in,clip=true,angle=-90]{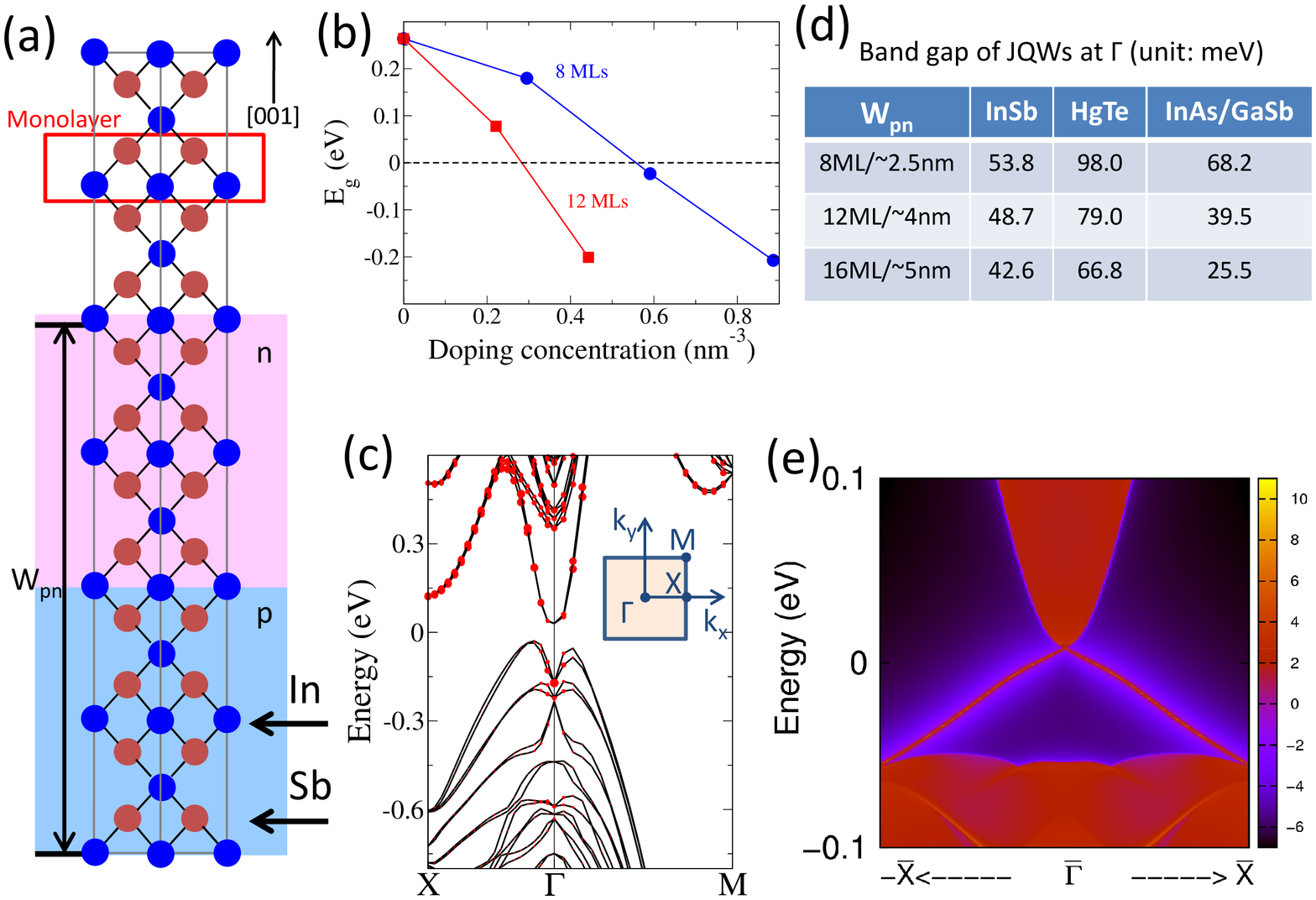}
\end{center}
\caption{(color online). (a) Schematic crystal structure of an InSb-based ($p$-InSb/$n$-InSb/InSb) JQW. The pink/blue background indicates the $p$-/$n$-region, and the residue is the intrinsic InSb. The growth direction is along the [001] direction. (b) $E_g$ at $\Gamma$ with different doping concentration for 8ML and 12ML InSb-based JQWs. Negative $E_g$ indicates a band inversion. (c) The band structure of the 12ML InSb-based JQW with the 0.46 nm$^{-3}$ carrier concentration. The red dots denote the projection of $s$ state of In in the $n$-region, which confirms the band inversion between the bottom of conduction and the top of valence bands. The schematic of Brillouin zone is plotted in the inset. (d)The calculated bulk band gaps for JQWs based on HgTe, InSb and InAs/GaSb with different $W_{pn}$. (e) The topologically protected edge states of 12ML InSb-based JQW with the QSH state.
 }\label{fig2}
\end{figure*}

{\bf Junction quantum wells.} A $p$-$n$ junction is the boundary or interface between $p$- and $n$-type doped semiconductors, usually inside a single material. In this work, we consider the limit where both the $p$-type and the $n$-type regions are narrow, and comparable in size to the interface region. We call such a device junction quantum well (JQW), which is depicted in Figs.~1(a) and (b). Once such a JQW realizes the QSH state, topologically protected edge states arise on its boundary, as marked by the red and blue curves in Fig.~1(a). Below we demonstrate, step by step, how to obtain QSH and QAH states in JQWs. First of all, before the charge transfer between the $p$- and $n$-regions, the chemical potential in the $n$-region is higher than that in the $p$-region, shown in Fig.~1(c). Secondly, after the charge transfer is turned on, electrons would move from the $n$- to the $p$-region due to the different chemical potentials, and an electric field is inevitably built up to stop the charge transfer to align chemical potentials of $p$- and $n$- regions, shown in Fig.~1(d). With the heavy doping (e.g. $\sim$0.1 nm$^{-3}$), a so-called band inversion between the conduction and valence bands can be consequentially obtained, schematically shown in Fig.~1(e). Such a band inversion is the key to obtain the QSH and QAH states,\cite{bernevig2006d}. If we define $E_g$ as the band gap at $\Gamma$, a negative $E_g$ is the indication of band inversion. Here we demonstrate the $s$-$p$-type band inversion, because most II-VI, III-V and IV semiconductors have the $s$-type conduction bands and $p$-type valence bands, for example, CdTe, InSb, Ge and so on. But the conclusion can be generalized to other kinds of band inversions, like the classification of three-dimensional topological insulators.\cite{zhang2012a} Thirdly, besides a band inversion, an insulating bulk state is also required for both QSH and QAH states, so these JQWs should be fully depleted, and the thickness of the depletion region (defined as W$_{pn}$) is known to be in nanoscale (around 10 nm). Importantly, the wavefunctions can well hybridize in nanoscale JQWs, and a bulk band gap, whose size depends on the strength of the hybridization, can be fully opened, shown in Fig.~1(f). Based on {\it ab-initio} calculations, the size of the bulk band gap can be as large as 0.1 eV (e.g. a HgTe-based JQW with the thickness $\sim$2.5 nm). In addition, the QSH state can be considered as two copies of QAH states which are coupled together by time reversal symmetry. After one QAH copy is removed by breaking time reversal symmetry through the magnetic doping, the transition from the QSH state to the QAH state can be realized.\cite{liu2008,Yu2010} Another interesting point is that a forward/backward bias on JQWs can weaken/strengthen the built-in electric field to tune the band inversion and realize a topological transition between the QSH state and the conventional insulating state. The physical picture described above can be applied to most JQWs of II-VI, III-V or IV semiconductors with a direct bulk band gap at $\Gamma$. Without loss of generality, we take InSb-based JQWs as an example to demonstrate how to design QSH and QAH states.

{\bf Methods.} {\it Ab-initio} calculations are carried out within the framework of the Perdew-Burke-Ernzerhof-type\cite{perdew1996} generalized gradient approximation of density functional theory\cite{hohenberg1964} by employing the BSTATE package\cite{fang2002} with the pseduopotential method and the Vienna ab-initio simulation package (VASP)\cite{kresse1993,kresse1999} with the projector augmented-wave method. The kinetic energy cutoff is fixed to be 450 eV. The spin-orbit coupling (SOC) is included. As is well known, the band gap of InSb is much underestimated by generalized gradient approximation of density functional theory, but a Coulomb repulsion $U$$_s$ (12 eV) on the $s$ orbital of In with the LDA+$U$ method \cite{dudarev1998} can be used to obtain the experimentally observed band gap.\cite{Boonchun2011} The LDA+$U$ method is also employed to do the supercell calculations of the magnetic doping with 4 eV $U_d$ on the $d$ orbitals of 3$d$ transition metals. We use the virtual crystal approximation\cite{xu2013a} to simulate the $p$-$n$ doping which is assumed to be homogeneous. A tight-binding method, based on maximally localized Wannier functions (MLWFs)\cite{marzari1997,souza2001}, is used to calculate the edge states of QSH and QAH states in JQWs, where the total Hamiltonian $H_{tot}$ is divided into the non-SOC part $H_{nosoc}$, the SOC part $H_{soc}$ and the exchange part $H_{ex}$. The $H_{nosoc}$ of an intrinsic InSb supercell with [001] as the normal direction is first constructed. The built-in electric field is introduced by a step potential to simulate the $p$-$n$ doping by fitting the band structure of {\it ab-initio} calculations. In addition, the form of $H_{soc}$ \cite{Jones2009} and $H_{ex}$ are written with atomic orbitals as basis because of the similarity between MLWFs and atomic orbitals in this case.

{\bf Crystal Structure.}
Bulk InSb has a rocksalt structure, the $Fm\overline{3}m$ space group (No. 225), with two independent atoms [one In at (0,0,0) and one Sb at ($\frac{1}{4}$,$\frac{1}{4}$,$\frac{1}{4}$)] in one unit cell. Its lattice constant is fixed to be the experimental value $6.47$\AA~.\cite{straumanis1965} Along the $z$ direction, we sequentially stack $p$-type, $n$-type InSb and intrinsic InSb (as the barrier) to construct a JQW, as schematically shown in Fig.~2(a). It has a binary axis (two-fold rotation symmetry) defined as the $z$ axis. Intrinsic InSb is taken as the barrier for InSb-based JQWs. We stress that not only the $z$ direction but also any other easy growth direction is suitable for our proposal of QSH and QAH states. Further, in order to insure an insulating bulk state, the carrier concentration of electrons and holes should be equal in $p$- and $n$- regions. For simplicity, we take the same carrier concentration and thickness for $p$- and $n$- regions.

{\bf Electronic structure.} In order to demonstrate the QSH state, {\it ab-initio} calculations are first carried out on InSb-based JQWs with two different $W_{pn}$, for example, 8 and 12 MLs, to confirm the band inversion. The band gap $E_g$ at $\Gamma$ is calculated with different carrier concentrations, shown in Fig.~2(b). We can see that the $E_g$ changes the sign from `$+$' to `$-$', indicating a band inversion, when the carrier concentration is around 0.6/0.3 nm$^{-3}$ for 8/12ML $W_{pn}$. Both the $W_{pn}$ and the carrier concentration are in reasonable range for experiments. The band structure of 12ML JQW with the carrier concentration 0.4 nm$^{-3}$ is shown in Fig.~2(c). The red dots represent the projection of In's $s$ states in the n region. We can see that partial In's $s$ state become occupied at $\Gamma$, and accordingly Sb's $p$ states (light hole states) switch to be unoccupied, which unambiguously confirms a $s$-$p$-type band inversion indicating the appearance of the QSH state. To further prove the QSH state, we employ the tight-binding method, based on MLWFs \cite{marzari1997,souza2001}, to calculate its edge states. The [010]/[100] direction is taken with an open/periodic boundary, respectively. In Fig.~2(e), topologically protected gapless edge states, including a single Dirac point, arise inside the bulk band gap. The bulk band gap is around 50 meV, which depends on the hybridization of the wavefunctions. In Fig.~2(d), we show the band gaps of some other typical JQWs of InSb, HgTe, and InAs/GaSb with different $W_{pn}$. We can see that the band gap is larger for thinner JQWs because of the larger hybridization. Especially, the 8ML HgTe-based JQWs gives a largest band gap $\sim$0.1 eV.

As mentioned above, time reversal symmetry needs to be broken to realize the transition from the QSH state to the QAH state by introducing the ferromagnetism. In most II-VI, III-V and IV semiconductors, the ferromagnetism can be obtained by magnetic doping, well known as diluted magnetic semiconductors (DMS)\cite{jungwirth2006,sato2010}, where most 3$d$ transition elements are widely-used dopants, for example, Mn. The element Mn usually exhibits the valence `+2' and a high spin state of $d$ orbitals. In order to see the $sp$-$d$ exchange splitting in details,\cite{jungwirth2006,sato2010} we construct a supercell with the formula MnIn$_{31}$Sb$_{32}$ (or In$_{0.96875}$Mn$_{0.03125}$Sb) for {\it ab-initio} calculations. The spin polarized DOS without SOC, shown in Fig.~3(a), indicate that the states around Fermi level most come from the In's $s$ and Sb's $p$ orbitals, and the states of Mn's $d$ orbitals stay far from Fermi level. The band structure is given in Fig.~3(b). The sign of the exchange splitting is opposite between the valence and conduction bands, which is important to realize the QAH state with a $s$-$p$ band inversion.\cite{liu2008} We also study the doping of other 3$d$ transition metals from Ti to Ni, and find that Fe and Cr are also suitable candidates for the QAH state, because for both Fe and Cr doping, similar to the Mn doping, the occupied $d$ states stay below the top of valence bands and the exchange splitting exhibits an opposite sign between valence and conduction bands, shown in Figs.~3(c-f). Commonly, free carriers are necessary to set up a ferromagnetism in DMS. However, it is an insulating state in InSb-based JQWs for our model, so seemingly it is impossible to get a ferromagnetic order within the same magnetic mechanism. But the fact is that, similar to the situation in Cr-doped Bi$_2$Te$_3$,\cite{Yu2010}  band edge singularities in the band-inversion regime play a role in inducing a ferromagnetic order. The detailed discussion can be found in a recent companion work by Wang et al.\cite{wang2013}

\begin{figure}[t]
\begin{center}
\includegraphics[width=5.8in,clip=true,angle=-90]{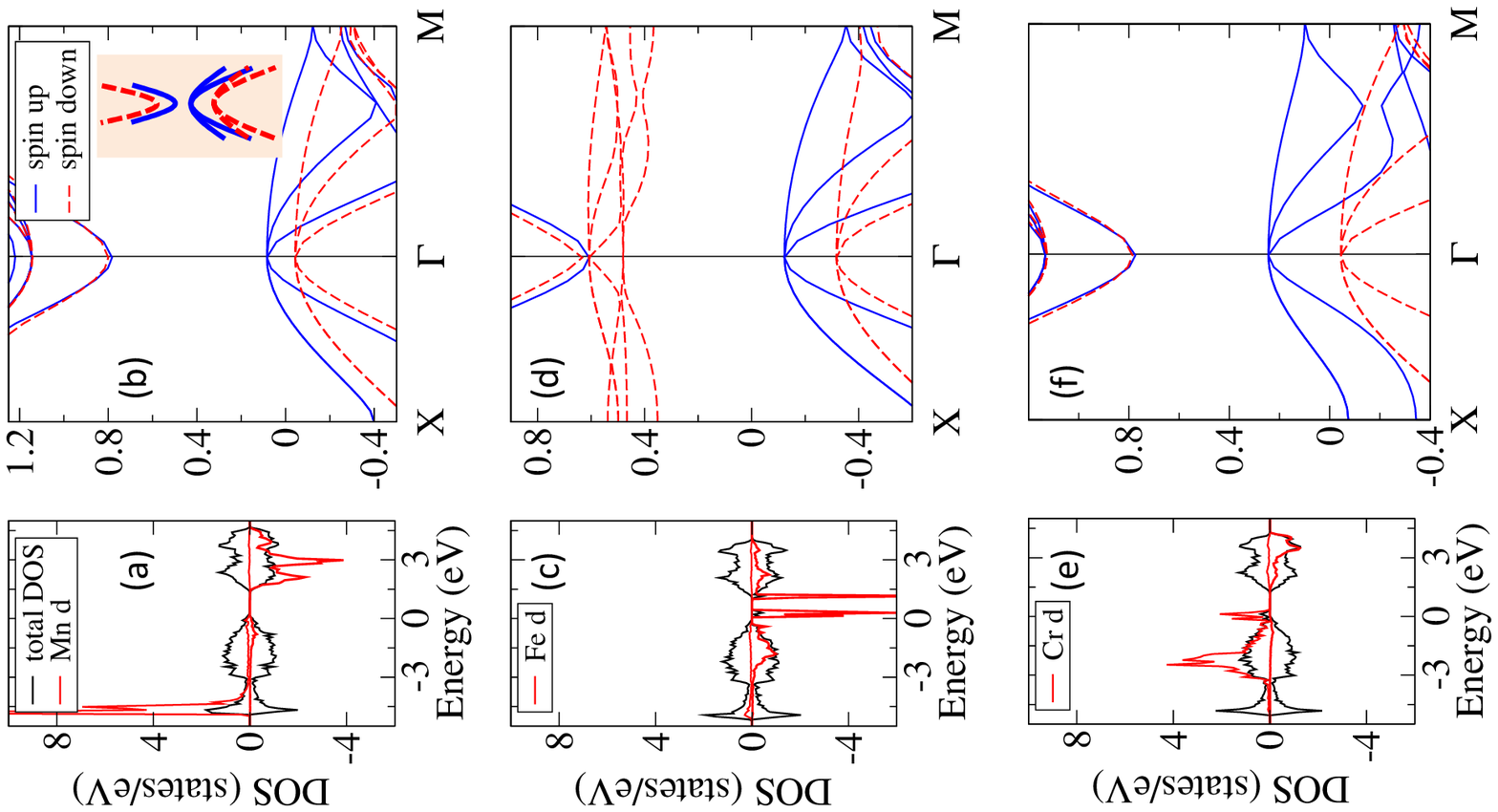}
\end{center}
\caption{(color online).(a) The spin polarized density of states (DOS) of Mn doped InSb by the formula (MnIn$_{31}$)Sb$_{32}$ (or In$_{0.96875}$Mn$_{0.03125}$Sb). The black/red curves are for the total/projected DOS. We stress that the total DOS is averaged to one unit cell. The positive/negative values of DOS are for spin up/down. Fermi level is fixed at 0 eV. (b) The band structure of (MnIn$_{31}$)Sb$_{32}$ without SOC. The solid blue/dashed red curves are for spin up/down bands. The schematic of the bands around Fermi level is shown in the inset. The exchange-splitting sign is opposite between the conduction and valence bands. (c-f) The spin polarized DOS and band structure of (\emph{M}In$_{31}$)Sb$_{32}$ with (\emph{M}=Fe, Cr).
 }\label{fig3}
\end{figure}

\begin{figure}[t]
\begin{center}
\includegraphics[width=3.2in,clip=true,angle=-90]{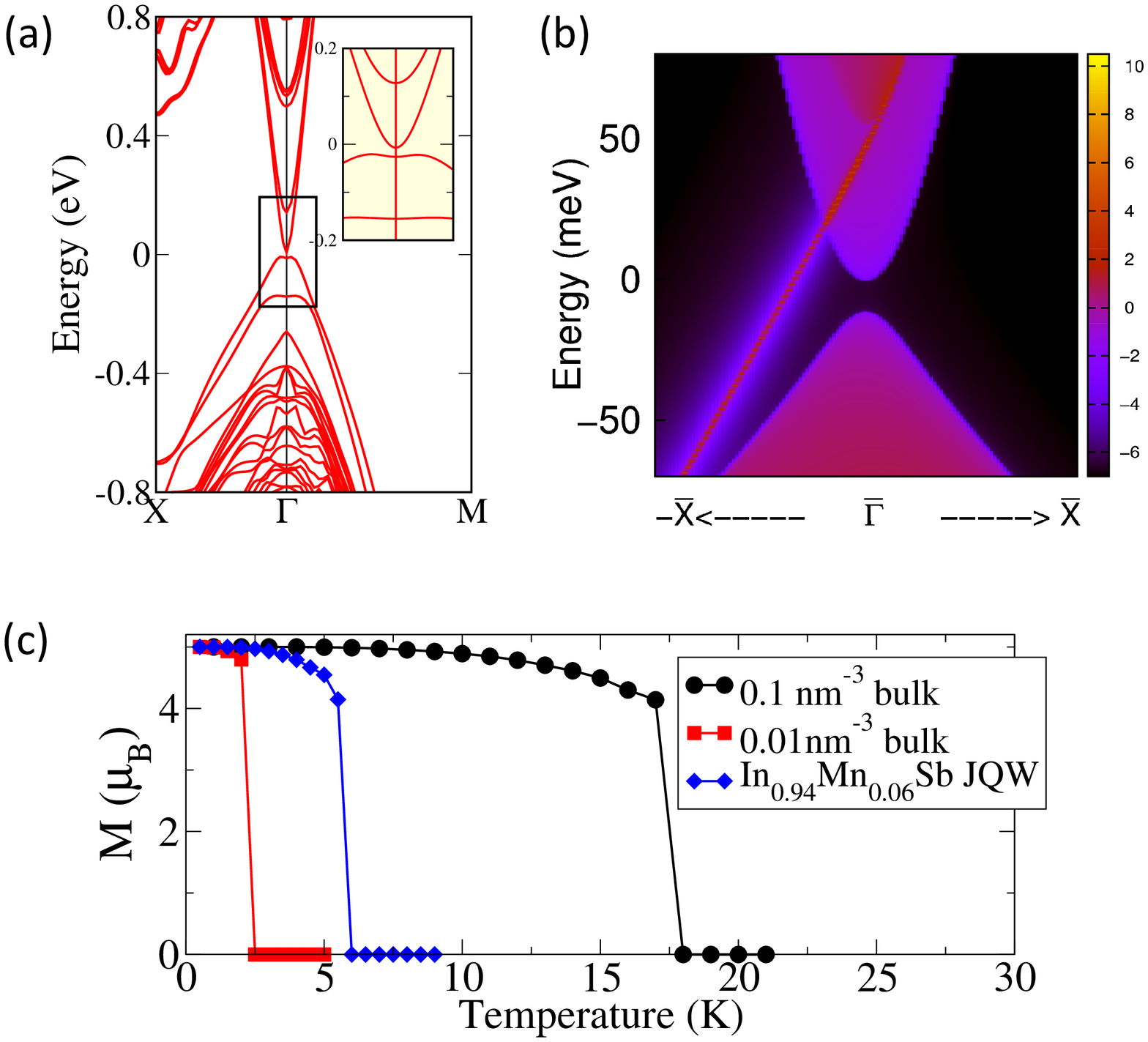}
\end{center}
\caption{(color online). The electronic structure of the QAH state. (a) The bulk band structure of a 12ML In$_{0.94}$Mn$_{0.06}$Sb JQW with the QAH state. The bands are zoomed-in in the inset.(b) The chiral edge states of the QAH state. (c) Magnetizing curves. The black circles/red squares are for the Mn-doped bulk InSb (In$_{0.95}$Mn$_{0.05}$Sb) with the hole concentration 0.1/0.01 nm$^{-3}$. The blue diamonds are for the 12ML In$_{0.94}$Mn$_{0.06}$Sb JQW.
 }\label{fig4}
\end{figure}

The QAH state is expected to be obtained by the combination of the built-in electric field and the $sp$-$d$ exchange interaction in InSb-based JQWs. In order to confirm this idea, we construct a tight-binding method of Mn-doped 12ML InSb-based JQW, based on MLWFs. The parameters of the exchange coupling are estimated based on the method of Dietl et al,\cite{Dietl2001} for exmple, $-$0.8/0.2 eV for $p$-$d$/$s$-$d$ exchange, and the Mn doping is taken as 6\%. Firstly, we estimate the Curie temperature $T_c$ based on a tight-binding Zener model\cite{Dietl2000,Dietl2001,jungwirth2006} under the mean-field frame. The perpendicular magnetization, as experimentally observed in Mn-doped InSb thin films\cite{abolfath2001,munekata1993,hayashi1997}, is fixed in our calculations. Figure 4(c) shows the calculated magnetizing curves, where we can see that the $T_c$ is 18K/2.5K for the bulk In$_{0.95}$Mn$_{0.05}$Sb with the 0.1/0.01 nm$^{-3}$ hole concentration, which agrees with other theoretical calculations and experiments,\cite{jungwirth2002} and 7K for the the 12ML InSb-based JQW. Secondly, edge states are calculated to confirm the QAH state, shown in Figs.~4(a) and (b). One chiral edge state really appears inside the band gap, and also we can see the band gap is around 5 meV which is dominated by the size of the $s$-$d$ exchange splitting in conduction bands.

{\bf Conclusion.} In summary, JQWs comprising II-VI, III-V or IV semiconductors with a direct band gap at $\Gamma$ are predicted to be a large class of new QSH and QAH materials, which offers broad possibilities for experiments to obtain the topological states. Especially most II-VI, III-V and IV semiconductors were well studied in the past several decades.\cite{yu1999} In addition, based on our calculations, the bulk band gap of QSH states can be as large as 0.1 eV which is much larger than that of HgTe/CdTe and type-II InAs/GaSb QWs. Also a bias can be used to switch QSH and conventional insulating states along with turning on or off the topological edge states, which provides a new functionality of JQWs for next-generation electronic devices. Further, DMS materials with II-VI, III-V or IV semiconductors were intensively studied by theories and experiments in the past, so the groundwork was well-established to obtain the QAH state in this kind of JQWs by the magnetic doping. QSH and QAH states would walk into us following this proposal.

\begin{acknowledgments}
We would like to thank Minghui Lu, Chaoxing Liu, Hong Yao and Zhong Wang for useful discussions. This work is supported by the US Department of Energy, Office of Basic Energy Sciences, Division of Materials Sciences and Engineering, under Contract No. DE-AC02-76SF00515, by the Defense Advanced Research Projects Agency Microsystems Technology Office, MesoDynamic Architecture Program (MESO) through the Contract No. N66001-11-1-4105 and by FAME, one of six centers of STARnet, a Semiconductor Research Corporation program sponsored by MARCO and DARPA.

\end{acknowledgments}
%


%

\end{document}